\documentclass{article}
\usepackage{spconf,amsmath,graphicx}
\usepackage{xcolor}

\usepackage{amsfonts}

\usepackage{booktabs}
\usepackage{multirow}
\usepackage{colortbl}
\usepackage{diagbox}
\usepackage{float}
\usepackage{makecell}
\usepackage{textcomp}
\usepackage{bbm}

\usepackage{amsmath}
\usepackage{graphicx}
\usepackage{pgfplots}
\usepackage{pgfplotstable}
\usepackage{caption}
\usepackage{subcaption}

\usepackage{tikz}
\usepackage{siunitx,etoolbox}

\usepackage{mathtools}

\definecolor{C0}{HTML}{1F77B4}
\definecolor{C1}{HTML}{FF7F0E}
\definecolor{C2}{HTML}{2CA02C}
\definecolor{C3}{HTML}{D62728}
\definecolor{C4}{HTML}{9467BD}
\definecolor{C5}{HTML}{8C564B}
\definecolor{C6}{HTML}{E377C2}
\definecolor{C7}{HTML}{7F7F7F}
\definecolor{C8}{HTML}{BCBD22}
\definecolor{C9}{HTML}{17BECF}

\title{Supervised Attention in Sequence-to-Sequence Models\\ for Speech Recognition}
\name{Gene-Ping Yang, Hao Tang}
\address{Centre for Speech Technology Research, School of Informatics, University of Edinburgh}

\begin{document}

\maketitle
\begin{abstract}
Attention mechanism in sequence-to-sequence models is
designed to model the alignments between acoustic features and output tokens in speech recognition.
However,
attention weights produced by models trained end to end do not always correspond well with actual alignments, and
several studies have further argued that attention weights might not even correspond well with the relevance attribution of frames.
Regardless, visual similarity between attention weights and alignments is widely used during training as an indicator of the models quality.
In this paper, we treat the correspondence between attention weights and alignments as a learning problem by imposing a supervised attention loss.
Experiments have shown significant improved performance, suggesting that learning the alignments well during training critically determines the performance of sequence-to-sequence models.
\end{abstract}

\section{Introduction}
\label{sec:intro}

Sequence-to-sequence models are extensively used in automatic speech recognition \cite{cho2015,bahdanau2016end,chan2016listen}.
The goal of the model is to take sequences of acoustic features as input and transform them into output sequences of phone, character or word tokens.
The model consists of three components: an encoder, a decoder and an attention mechanism
that bridges the information between the encoder and the decoder.
Given a sequence of acoustic features, the encoder computes a sequence of hidden vectors, each of which is assigned a weight by the attention mechanism.
The decoder takes the weighted sum of the hidden vectors and its own hidden vector to generate the output token.

The attention mechanism plays an important role in sequence-to-sequence models as it allows the models to weight the encoded input acoustic features at different decoding steps.
The attention mechanism is designed to model alignments, the natural correspondence between acoustic features and output tokens in speech recognition \cite{Bahdanau2015NeuralMT, luong-etal-2015-effective, cho2015}.
However, given how little attention is typically constrained during end-to-end training, the attention weights do not always behave like alignments \cite{serrano-smith-2019-attention, jainwallace2019}.
Several studies have shown that different attention weights can lead to similar outputs and the attention weights might not correspond well with the relevance attribution of frames \cite{jainwallace2019}.
This paper studies this problem and attempts to understand if this is a problem at all.

It is a common belief that how well attention weights correspond to alignments is a good indicator of model performance \cite{kim2017joint}.
In fact, many have used this heuristic during model training or for hyperparameter tuning \cite{kim2017joint, 7472655}.
In this paper, we frame this common belief as a scientific question: \emph{Does having a strong correspondence between attention weights and alignments lead to better performance? Is it necessary to have a strong correspondence to achieve good performance?}
We introduce a distance between attention weights and alignments to measure the correspondence and provide supervision signal to the attention during training.
Finding the correspondence between attention weights and alignments becomes a learning problem, and whether we can faithfully interpret attention becomes a question of generalization.

Owing to the common belief that model performance is better when attention behaves like alignments, many approaches have been proposed to tighten the correspondence~\cite{kim2017joint, ChiuR18, nguyen-etal-2020-differentiable, 8682224}.
It is in fact a common practice to supervise attention against an alignment where the duration is equally divided among the output tokens \cite{Hannun2019}. This alignment is obviously incorrect, but it serves to encourage monotonicity, especially early during training. The supervision is then turned off after a few epochs \cite{Hannun2019}.
These approaches assume that the correspondence between attention and alignments critically depends on how well the models discover alignments \emph{early} during training.
If the assumption is correct, we would observe a faster convergence if we supply the supervision signal with better alignments.
In this paper, we further study this assumption with curriculum training \cite{bengio2009curriculum}, turning off the supervision midway through training.

Our results support the common believes. In addition, supervised attention and curriculum training both significantly improve the performance of sequence-to-sequence models.

\section{Proposed Approach}
\label{sec:format}

We first review the formulation of sequence-to-sequence models in this section, and will
discuss the options of converting alignments into attention weights and the loss of supervised attention.

\subsection{Sequence-to-Sequence Models}

    Given an input sequence $x_1, x_2, ..,. x_T$ of length $T$ and target sequence $y_1, y_2, ..., y_K$ of length $K$,
    the conditional probability of the output sequence given the input sequence can be factorized as
    \begin{equation}
        P(y_{1:K}| x_{1:T}) = \prod_{k=1}^K P(y_k | x_{1:T}, y_{1:k-1}),
    \end{equation}
    where $x_{1:t}$ stands for $x_1, \dots, x_t$ and similarly for $y_{1:k}$. 
    A sequence-to-sequence model defines the probability distribution $P(y_k | x_{1:T}, y_{1:k-1})$, i.e., the probability of the next output token given all the inputs and the past output tokens.
    
    First, the encoder transforms the input sequence into a sequence of hidden vectors $h_1, h_2, ..., h_{T'}$ of length $T'$.
    \begin{equation}
        h_{1:T'} = \text{Enc}(x_{1:T})
    \end{equation}
    Note that $T'$ and $T$ do not necessarily need to be the same. 
    It is common to subsample the hidden vectors, for example, by a factor of four,
    \cite{chan2016listen, vanhoucke2013multiframe, miao2016simplifying}.
    In that case, $T' = \lfloor T/4 \rfloor$.
    For the $k$-th output token, the decoder first encodes the past output tokens into a vector
    \begin{equation}
       d_k = \text{Dec}(y_{1:k-1}).
    \end{equation}
    The attention weights represented as a vector $\alpha_k$ are defined as
    \begin{equation}
       \alpha_{k, t} = \frac{\exp (d_k^\top h_t)}{\sum_{j=1}^{T'} \exp (d_k^\top h_j)}.
    \end{equation}
    In other words, each attention weight is computed based on the similarity of a hidden vector $h_t$ and $d_k$, followed by a softmax, which constrains the weights to sum to one.
    The attention weights are then used to compute a context vector
    \begin{equation}
       c_k = \sum_{j=1}^{T'} \alpha_{k, j} h_j.
    \end{equation}
    Finally, the context vector $c_k$ and the decoder hidden vector $d_k$ are used for prediction, i.e.,
    \begin{equation}
        P(y_{k, i} | x_{1:T}, y_{1:k-1}) = \frac{\exp({\phi_{k}[i]})}{\sum_{v=1}^V \exp({\phi_{k}[v]})}
    \end{equation}
    where $\phi_k = W \begin{bmatrix}c_k \\ d_k\end{bmatrix}$ and $V$ is the size of the output token set.
    
    To train the model, for each pair of $x_{1:T}$ and $y_{1:K}$ in the training set, we optimize the loss function
    \begin{equation}
        L_{\text{ce}} = -\ln P(y_{1:K} | x_{1:T}),
    \end{equation}
    which is commonly known as teacher forcing \cite{williams1989learning, lamb2016}. 
    
\subsection{Supervised Attention}

    The attention weights $\alpha$ can be represented as a matrix, where each element $\alpha_{k, i}$ is the similarity score of the encoder hidden vector $h_i$ and the decoder hidden vector $d_k$.

    An alignment is a sequence of 3-tuples $(s_1, e_1, y_1)$, $\dots$, $(s_K, e_K, y_K)$, where $s_k$ is the start time and $e_k$ is the end time, indicating the time span of a segment $y_k$. 
    In this paper, we convert an alignment into the attention weights $\alpha^{*}$ by assigning uniform weights to the vectors within the corresponding start time and end time.
    Formally,
    \begin{equation}
        \alpha_{k, t}^* = \frac{1}{e_k - s_k} \mathbbm{1}_{s_k \leq t < e_k},
    \end{equation}
    where $\mathbbm{1}_c$ is 1 if $c$ is true, and 0 otherwise.
    When subsampling is involved, say by a factor of $r$, we simply sum the corresponding weights together. 
    Specifically, the attention weight at $k, t'$ after subsampling is
    $\sum_{t=r(t'-1)}^{rt'} \alpha_{k,t}^{*}$.
    
    Although it is intuitive to assign uniform weights within a segment, peaky attention weights is often observed where only a few encoder hidden vectors are considered during decoding.
    Instead of assigning uniform weights within a segment, we explore alternatives such as assigning a point mass at the first, center, and last frame of a segment.
    For example, $\alpha_{k, t}^* = \mathbbm{1}_{t = (s_k+e_k) / 2}$ if a point mass is assigned at the center of a segment.

    We can also write the alignment where the duration is equally divided among segments using our notation.
    With input sequence of length $T$ and output sequence of length $K$,
    \begin{equation}
        \alpha_{k, t}^* = \frac{1}{d} \mathbbm{1}_{kd \leq t < (k+1)d},
    \end{equation}
    where $d=T/K$ is the average duration of each segment.
    
    Once an alignment is converted into attention weights, we use the Frobenius norm to measure the distance between $\alpha$ and $\alpha^{*}$, and introduces an additional loss
    \begin{equation}
        L_{\text{attn}} = \|\alpha^* - \alpha\|_F^2
    \end{equation}
    as additional objective to train the model.
    The overall loss function is
    \begin{equation}
        L = L_{\text{ce}} + \gamma L_{\text{attn}},
    \end{equation}
    where $\gamma$ is a hyperparameter.
    
    The hyperparameter $\gamma$ can change during the course of training.
    We explore two cases, one where $\gamma$ is set to a positive value throughout, and the other where we change $\gamma$ from a fixed positive value to zero, disabling supervised attention, halfway through training.

\section{Related Work}

    Supervised attention against evenly divided alignments has been used for encouraging monotonicity early during training.
    This is a common practice, though seldom mentioned in studies, except in \cite{Hannun2019}.
    Supervised attention against forced alignments has been explored in speech recognition, with various distances to measure the correspondence between attention and alignments, and with various representation of alignments~\cite{superattn2020}.
    Our focus is on the analysis of how the correspondence relates to model performance, and of how various design decisions change the behavior of the model.
    We also make connections to the common believes, common practices, and curriculum training \cite{bengio2009curriculum}.
    
    More broadly, supervised attention has been applied in text classification \cite{zhang2016rationale} and neural machine translation~\cite{mi-etal-2016-supervised, liu-etal-2016-neural, dou2021attention}.
    These prior work also shares similar concerns and believes that attention mechanism is too flexible under end-to-end training and prone to having unintuitive and unfaithful interpretation.
    Similar ideas has also been explored in text-to-speech synthesis \cite{zhu2019pag, dou2020attention}.
    In addition to applying supervised attention, the study \cite{zhu2019pag} also showcases that tuning models based on the outcome of attention has become a common practice, and further strengthens the common belief that tight correspondence between attention and alignments implies good model performance.

    There is an ongoing debate whether attention can be interpreted as alignments \cite{jainwallace2019,serrano-smith-2019-attention, wiegreffe-pinter-2019-attention}.
    If attention mechanism were trustworthy, it would not have spawned the variants \cite{ChiuR18,8682224, nguyen-etal-2020-differentiable} and other techniques \cite{Hannun2019,kim2017joint} to improve the mechanism.
    Regardless, attention has been used for word segmentation \cite{boito2019, Godard2018}, though the robustness of this approach has been questioned \cite{sanabria2021difficulty}.
    Our approach sidesteps this problem by turning the correspondence between attention and alignments as a learning problem.
    Whether we can obtain faithful interpretation becomes a question of generalization.
    
\section{Experiments}

    We design phonetic recognition experiments on the Wall Street Journal dataset to understand how the correspondence between attention weights and alignments relates to model performance.
    The training set \texttt{si284} is split into training and development set with a ratio of $9:1$. We map words into their canonical pronunciations according to \texttt{cmudict}.
    The label set, with stress markers removed, includes 39 phones and three special tokens for silence (\texttt{sil}) and noise (\texttt{spn} and \texttt{nsn}).\footnote{Phonetic recognition on WSJ has been explored in \cite{miao2015eesen, chen20m_interspeech}, but there are no consensus as to what label sets to use. None of the results to be presented are directly comparable to prior work, hence the additional baseline numbers provided later in this section.}
    Forced alignments produced by a speaker-adaptive GMM-HMM are used for supervised attention.
    We tune the hyperparameters on the development set and report the final numbers on dev93 and eval92. We use 40-dimensional log Mel spectrograms as input, without the first and second-order derivatives. Global (instead of speaker-dependent) mean and variance normalization is applied to the input features.
    
    Our model architecture follows \cite{kim2017joint}. 
    The encoder is a 4-layer bidirectional LSTM, with 320 cells in each direction and each layer.
    Subsampling of factor two is done twice, one after the $2^{\text{nd}}$ layer and another after the $3^{\text{rd}}$ layer of the encoder, resulting in $1/4$ of the original frame rate. 
    The decoder consists of a phone embedding layer, a 1-layer unidirectional LSTM, and 2 fully connected layers after the LSTM.
    We use teacher forcing throughout training and greedy decoding during testing.
    We tune the value of $\gamma$ and the dropout rate on the validation set and choose $\gamma = 0.5$ and a dropout rate of 0.4 throughout all experiments.
    Following \cite{kim2017joint}, all models are trained with Adadelta \cite{zeiler2012adadelta} for 30 epochs.

    \begin{table}
        \caption{Phone error rates (\%) on dev93 and eval92. s2s denotes the cross entropy loss.
        The loss uni-attn puts uniform weights within a segment.
        The losses f-attn, c-attn, and l-attn put a point mass at the first, center, and last frame of a segment, respectively.
        The loss even-attn divides the duration evenly for all segments.
        The last row is a CTC baseline.
        }
        \centering
        \begin{tabular}{lccc}
            \toprule
            Models & dev93 & eval92 \\
            \midrule
            s2s & 24.8 & 20.9 \\
            s2s + uni-attn & 12.1 & 8.7 \\
            \midrule
            s2s + dropout  & 15.3 & 11.8 \\
            s2s + dropout + uni-attn & \textbf{10.4} & \textbf{7.7} \\  
            s2s + dropout + f-attn & 13.5 & 9.7 \\
            s2s + dropout + c-attn   & 11.3 & 9.4 \\
            s2s + dropout + l-attn  & 12.4 & 9.5 \\
            s2s + dropout + even-attn  & 11.9 & 8.7 \\
            \midrule
            ctc + dropout & 9.5 & 6.7 \\
            \bottomrule
        \end{tabular}
        \label{tbl:sup-attn}
    \end{table}
    
    \subsection{Supervised attention improves performance}
    
    In Table \ref{tbl:sup-attn}, we show the performance of supervised attention.
    Adding dropout produces a stronger baseline.
    However, merely using supervised attention without dropout already outperforms the stronger baseline, and the improvement is still observed with dropout.
    Improvements are also observed regardless of how forced alignments are represented, with putting uniform weights within segments performs the best.
    CTC is included here for completeness. It is not surprising that CTC outperforms all sequence-to-sequence models on this task, a result consistent with prior work \cite{graves2013speech,cho2015}. 

    As an aside, we successfully train models that produces peaky attention while performing well (f-attn, c-attn, l-attn in Table~\ref{tbl:sup-attn}). This refutes the common belief that peaky attention is generally undesirable.
    What is surprising is that using incorrect alignments for supervised attention (even-attn in Table~\ref{tbl:sup-attn}) can still lead to an improvement.
    This gives a counter example and refutes the hypothesis that a tight correspondence between attention and alignments is necessary for good model performance.

    \subsection{Supervised attention improves convergence}
    
\begin{figure}[t]
    {\centering
    \begin{subfigure}[b]{0.23\textwidth}
    \centering
    \includegraphics[width=\textwidth]{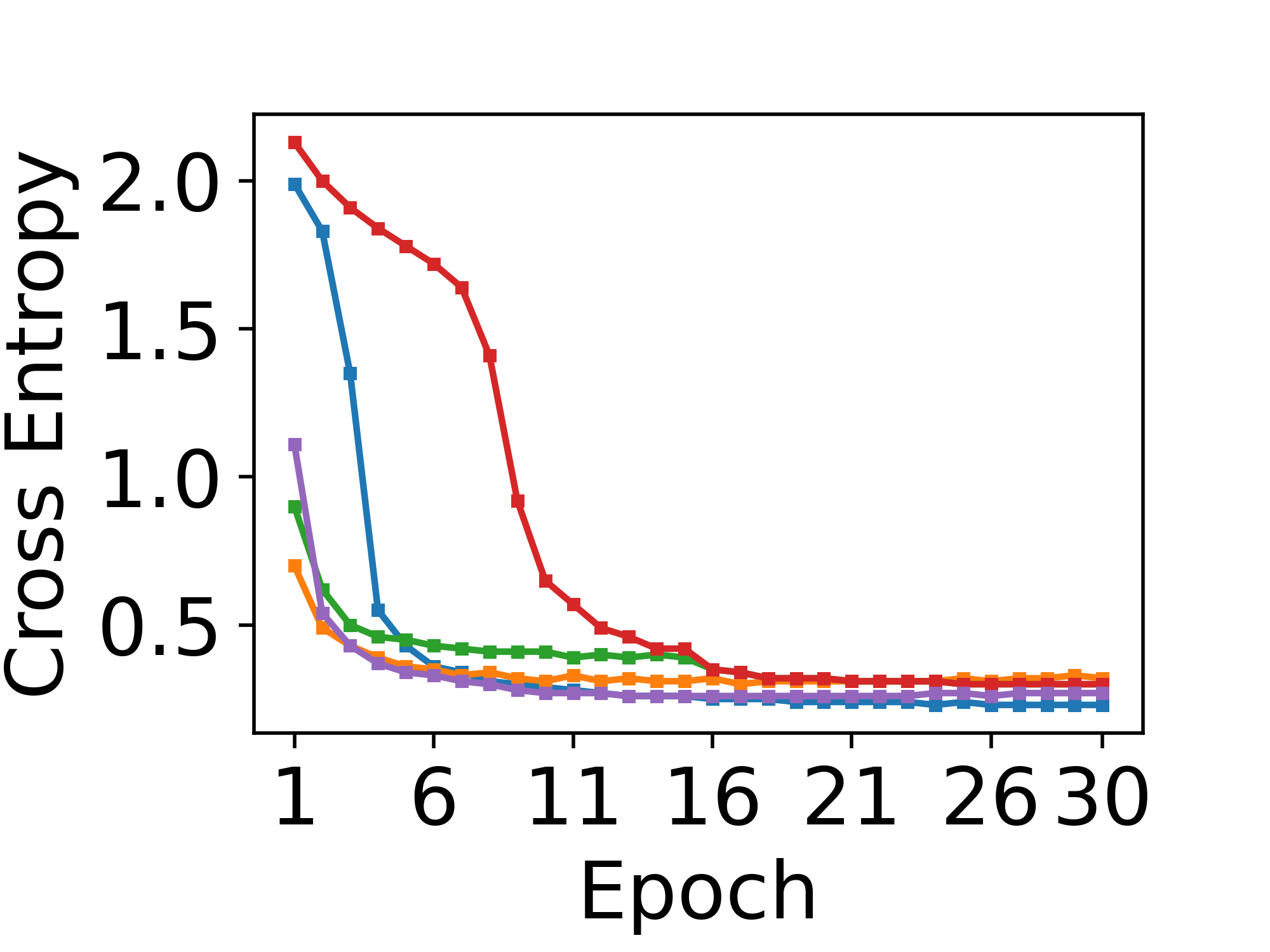}
    \caption{cross-entropy loss}
    \end{subfigure}
    \begin{subfigure}[b]{0.23\textwidth}
    \centering
    \includegraphics[width=\textwidth]{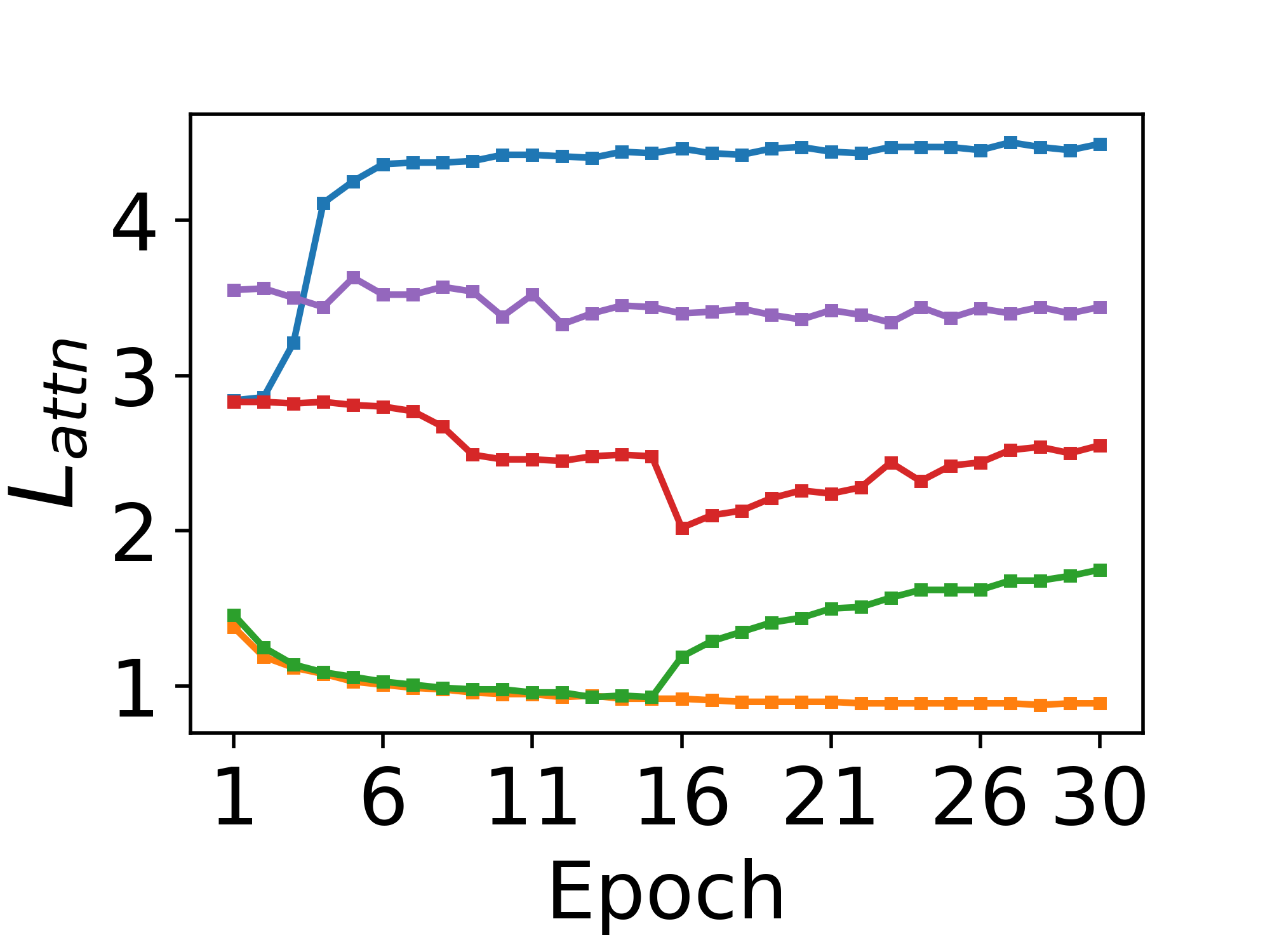}
    \caption{supervised attention loss}
    \end{subfigure}
    }
    \tikz[baseline=-3pt] \draw[C0] (0, 0) edge node[C0, fill, rectangle, inner sep=1pt] {} (0.5, 0); s2s
    \tikz[baseline=-3pt] \draw[C1] (0, 0) edge node[C1, fill, rectangle, inner sep=1pt] {} (0.5, 0); s2s + uni-attn
    \tikz[baseline=-3pt] \draw[C2] (0, 0) edge node[C2, fill, rectangle, inner sep=1pt] {} (0.5, 0); s2s + uni-attn + CT
    \tikz[baseline=-3pt] \draw[C3] (0, 0) edge node[C3, fill, rectangle, inner sep=1pt] {} (0.5, 0); s2s + even-attn + CT
    \tikz[baseline=-3pt] \draw[C4] (0, 0) edge node[C4, fill, rectangle, inner sep=1pt] {} (0.5, 0); s2s + ctc
    \caption{The cross entropy loss and the supervised attention loss on the development set along the course of training. The attention loss is computed against the forced alignments where attention weights are uniform within segments.
    We do not expect the losses other than uni-attn to be low, but the ones of uni-attn should, if they generalize well.}
    \label{curves}
    \end{figure}
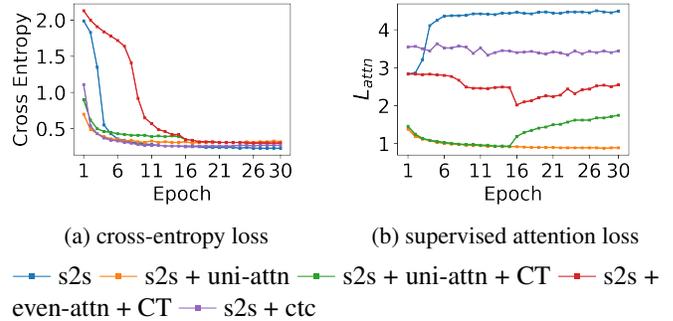

    In Fig \ref{curves}, we show (a) the cross-entropy loss and (b) the distance between predicted attention weights and forced alignments on the development set along the course of training.
    Models using forced alignments (uni-attn) converge faster in terms of cross entropy, suggesting that supervised attention significantly eases the learning of the task.
    However, with the attention weights where duration of segments is evenly divided (even-attn), the cross entropy converges slower, worse than the baseline.
    We also show that the model trained with attention that puts uniform weights within segments generalize well in terms of the supervised attention loss, producing a model that adheres to faithful interpretation.

    \begin{table}
        \caption{Phone error rates (\%) with curriculum training (CT) and the CTC loss.
        The loss uni-attn puts uniform weights within a segment, while the loss even-attn divides the duration evenly for all segments.}
        \centering
        \begin{tabular}{lccc}
            \toprule
            Models & dev93 & eval92 \\
            \midrule
            s2s + dropout + uni-attn & 10.4 & 7.7 \\  
            s2s + dropout + uni-attn + CT & 9.5 & 7.3 \\
            s2s + dropout + even-attn  & 11.9 & 8.7 \\
            s2s + dropout + even-attn + CT & 10.8 & 8.0 \\
            \midrule
            s2s + dropout + ctc & 12.7 & 7.8 \\
            s2s + dropout + ctc + uni-attn & \textbf{9.4} & \textbf{6.6} \\
            s2s + dropout + ctc + uni-attn + CT & 10.3 & 6.5 \\
            \bottomrule
        \end{tabular}
        \label{tbl:ct-ctc}
    \end{table}

    \subsection{Curriculum training further improves performance}
    
    To  study whether supervised attention continues to have an impact beyond the early stage of training,
    we adopt curriculum training,
    enabling supervised attention for the first 15 epochs
    and disabling it for the rest of the 15 epochs.
    We study the best performing setting where attention weights are put uniformly within segments (uni-attn) and where the duration is evenly divided for all segments (even-attn).
    The latter (even-attn) is meant to reproduce the common practice used in \cite{Hannun2019}.
    The results are shown in the top half of Table \ref{tbl:ct-ctc}.
    Both cases show further performance gain when paired with curriculum training.
    This confirms the usefulness of the common practice, while using forced alignments in supervised attention still performs better than the others. 
    
    In Fig \ref{curves} (b), we show how the supervised attention loss changes during the course of training.
    The loss first decreases but starts to go up once the supervised attention loss is turned off, ending at a loss value higher then the initial loss value.
    Given that the model performs better in the end, this result suggests that supervised attention might be too strong a constraint.
    This result also suggests that the model might have an inherent preference over other shapes of attention weights than having uniform weights within segments.
    
    \subsection{Supervised attention and CTC are complementary}
    
    CTC has also been used as a loss to tighten the correspondence between attention and alignments, because CTC only considers monotonic alignments \cite{kim2017joint}.
    We first reproduce the joint training approach proposed in \cite{kim2017joint}, and include supervised attention.
    The results are shown in the bottom half of Table \ref{tbl:ct-ctc}.
    Supervised attention with forced alignments is still better than the joint training approach.
    In addition, adding supervised attention to the joint training approach further reduces the phone error rates, suggesting that the two losses might be complementary.
    We do not get any gain by adding curriculum learning on top of the combination of the two.
    Tuning all combination becomes tedious and does not show any further insights even if we obtain a better performance.

\section{Conclusion}

    Supervised attention provides a simple yet effective step towards understanding the attention mechanism in sequence-to-sequence models.
    In addition, faster convergence and better performance have been observed.
    The experiments in this paper assume we have access to accurate forced alignments, a requirement that can be overly stringent.
    Nevertheless, obtaining accurate forced alignments is a simpler task than automatic speech recognition. 
    Several obvious extensions have been proposed, for example, in \cite{superattn2020}.
    The next step is to move beyond phonetic recognition, though it is nontrivial to align characters or word pieces.
    Syllables and words are better candidates, where boundaries are more consistent.
    In terms of analysis, more effort should be put into what information is attended and why other frames are \emph{not} attended, especially when the attention is peaky.

\bibliographystyle{IEEEbib}
\bibliography{strings,refs}

\begin{thebibliography}{10}

\bibitem{cho2015}
Jan Chorowski, Dzmitry Bahdanau, Dmitriy Serdyuk, Kyunghyun Cho, and Yoshua
  Bengio,
\newblock ``Attention-based models for speech recognition,''
\newblock in {\em Advances in Neural Information Processing Systems}, 2015.

\bibitem{bahdanau2016end}
Dzmitry Bahdanau, Jan Chorowski, Dmitriy Serdyuk, Philemon Brakel, and Yoshua
  Bengio,
\newblock ``End-to-end attention-based large vocabulary speech recognition,''
\newblock in {\em ICASSP}, 2016.

\bibitem{chan2016listen}
William Chan, Navdeep Jaitly, Quoc Le, and Oriol Vinyals,
\newblock ``Listen, attend and spell: A neural network for large vocabulary
  conversational speech recognition,''
\newblock in {\em ICASSP}, 2016.

\bibitem{Bahdanau2015NeuralMT}
Dzmitry Bahdanau, Kyunghyun Cho, and Yoshua Bengio,
\newblock ``Neural machine translation by jointly learning to align and
  translate,''
\newblock in {\em ICLR}, 2015.

\bibitem{luong-etal-2015-effective}
Thang Luong, Hieu Pham, and Christopher~D. Manning,
\newblock ``Effective approaches to attention-based neural machine
  translation,''
\newblock in {\em EMNLP}, 2015.

\bibitem{serrano-smith-2019-attention}
Sofia Serrano and Noah~A. Smith,
\newblock ``Is attention interpretable?,''
\newblock in {\em ACL}, 2019.

\bibitem{jainwallace2019}
Sarthak Jain and Byron~C. Wallace,
\newblock ``Attention is not explanation,''
\newblock in {\em NAACL}, 2019.

\bibitem{kim2017joint}
Suyoun Kim, Takaaki Hori, and Shinji Watanabe,
\newblock ``Joint {CTC}-attention based end-to-end speech recognition using
  multi-task learning,''
\newblock in {\em ICASSP}, 2017.

\bibitem{7472655}
Gustav~Eje Henter, Srikanth Ronanki, Oliver Watts, Mirjam Wester, Zhizheng Wu,
  and Simon King,
\newblock ``Robust {TTS} duration modelling using {DNN}s,''
\newblock in {\em ICASSP}, 2016.

\bibitem{ChiuR18}
{Chung-Cheng} Chiu and Colin Raffel,
\newblock ``Monotonic chunkwise attention,''
\newblock in {\em ICLR}, 2018.

\bibitem{nguyen-etal-2020-differentiable}
Thanh-Tung Nguyen, Xuan-Phi Nguyen, Shafiq Joty, and Xiaoli Li,
\newblock ``Differentiable window for dynamic local attention,''
\newblock in {\em ACL}, 2020.

\bibitem{8682224}
Shucong Zhang, Erfan Loweimi, Peter Bell, and Steve Renals,
\newblock ``Windowed attention mechanisms for speech recognition,''
\newblock in {\em ICASSP}, 2019.

\bibitem{Hannun2019}
Awni Hannun, Ann Lee, Qiantong Xu, and Ronan Collobert,
\newblock ``Sequence-to-sequence speech recognition with time-depth separable
  convolutions,''
\newblock in {\em Interspeech}, 2019.

\bibitem{bengio2009curriculum}
Yoshua Bengio, J{\'e}r{\^o}me Louradour, Ronan Collobert, and Jason Weston,
\newblock ``Curriculum learning,''
\newblock in {\em ICML}, 2009.

\bibitem{vanhoucke2013multiframe}
Vincent Vanhoucke, Matthieu Devin, and Georg Heigold,
\newblock ``Multiframe deep neural networks for acoustic modeling,''
\newblock in {\em ICASSP}, 2013.

\bibitem{miao2016simplifying}
Yajie Miao, Jinyu Li, Yongqiang Wang, Shi-Xiong Zhang, and Yifan Gong,
\newblock ``Simplifying long short-term memory acoustic models for fast
  training and decoding,''
\newblock in {\em ICASSP}, 2016.

\bibitem{williams1989learning}
Ronald~J Williams and David Zipser,
\newblock ``A learning algorithm for continually running fully recurrent neural
  networks,''
\newblock {\em Neural computation}, vol. 1, no. 2, pp. 270--280, 1989.

\bibitem{lamb2016}
Alex Lamb, Anirudh Goyal, Ying Zhang, Saizheng Zhang, Aaron Courville, and
  Yoshua Bengio,
\newblock ``Professor forcing: a new algorithm for training recurrent
  networks,''
\newblock in {\em Advances in Neural Information Processing Systems}, 2016.

\bibitem{superattn2020}
Shreekantha Nadig, Sumit Chakraborty, Anuj Shah, Chaitanay Sharma,
  V.~Ramasubramanian, and Sachit Rao,
\newblock ``Jointly learning to align and transcribe using attention-based
  alignment and uncertainty-to-weigh losses,''
\newblock in {\em International Conference on Signal Processing and
  Communications}, 2020.

\bibitem{zhang2016rationale}
Ye~Zhang, Iain Marshall, and Byron~C. Wallace,
\newblock ``Rationale-augmented convolutional neural networks for text
  classification,''
\newblock in {\em EMNLP}, 2016.

\bibitem{mi-etal-2016-supervised}
Haitao Mi, Zhiguo Wang, and Abe Ittycheriah,
\newblock ``Supervised attentions for neural machine translation,''
\newblock in {\em EMNLP}, 2016.

\bibitem{liu-etal-2016-neural}
Lemao Liu, Masao Utiyama, Andrew Finch, and Eiichiro Sumita,
\newblock ``Neural machine translation with supervised attention,''
\newblock in {\em International Conference on Computational Linguistics}, 2016.

\bibitem{dou2021attention}
Qingyun Dou, Yiting Lu, Potsawee Manakul, Xixin Wu, and Mark~JF Gales,
\newblock ``Attention forcing for machine translation,''
\newblock {\em arXiv preprint arXiv:2104.01264}, 2021.

\bibitem{zhu2019pag}
Xiaolian Zhu, Yuchao Zhang, Shan Yang, Liumeng Xue, and Lei Xie,
\newblock ``Pre-alignment guided attention for improving training efficiency
  and model stability in end-to-end speech synthesis,''
\newblock {\em IEEE Access}, 2019.

\bibitem{dou2020attention}
Qingyun Dou, Joshua Efiong, and Mark~JF Gales,
\newblock ``Attention forcing for speech synthesis,''
\newblock in {\em Interspeech}, 2020.

\bibitem{wiegreffe-pinter-2019-attention}
Sarah Wiegreffe and Yuval Pinter,
\newblock ``Attention is not not explanation,''
\newblock in {\em EMNLP}, 2019.

\bibitem{boito2019}
Marcely~Zanon Boito, Aline Villavicencio, and Laurent Besacier,
\newblock ``Empirical evaluation of sequence-to-sequence models for word
  discovery in low-resource settings,''
\newblock in {\em Interspeech}, 2019.

\bibitem{Godard2018}
Pierre Godard, Marcely~Zanon Boito, Lucas Ondel, Alexandre Berard, François
  Yvon, Aline Villavicencio, and Laurent Besacier,
\newblock ``Unsupervised word segmentation from speech with attention,''
\newblock in {\em Interspeech}, 2018.

\bibitem{sanabria2021difficulty}
Ramon Sanabria, Hao Tang, and Sharon Goldwater,
\newblock ``On the difficulty of segmenting words with attention,''
\newblock {\em arXiv preprint arXiv:2109.10107}, 2021.

\bibitem{miao2015eesen}
Yajie Miao, Mohammad Gowayyed, and Florian Metze,
\newblock ``{EESEN}: End-to-end speech recognition using deep {RNN} models and
  {WFST}-based decoding,''
\newblock in {\em ASRU}, 2015.

\bibitem{chen20m_interspeech}
Yang Chen, Weiran Wang, and Chao Wang,
\newblock ``Semi-supervised {ASR} by end-to-end self-training,''
\newblock in {\em Interspeech}, 2020.

\bibitem{zeiler2012adadelta}
Matthew~D Zeiler,
\newblock ``Adadelta: an adaptive learning rate method,''
\newblock {\em arXiv preprint arXiv:1212.5701}, 2012.

\bibitem{graves2013speech}
Alex Graves, {Abdel-rahman} Mohamed, and Geoffrey Hinton,
\newblock ``Speech recognition with deep recurrent neural networks,''
\newblock in {\em ICASSP}, 2013.

\end{thebibliography}

\end{document}